\documentclass[conference]{IEEEtran}
\IEEEoverridecommandlockouts

\usepackage{cite}
\usepackage{amsmath,amssymb,amsfonts}
\usepackage{algorithmic}
\usepackage{enumerate}
\usepackage{graphicx}
\usepackage{textcomp}
\usepackage{xcolor}
\usepackage{enumitem}
\usepackage{url}
\usepackage{multirow}
\usepackage{booktabs}
\usepackage{tabularx}
\def\BibTeX{{\rm B\kern-.05em{\sc i\kern-.025em b}\kern-.08em
    T\kern-.1667em\lower.7ex\hbox{E}\kern-.125emX}}
\usepackage{tcolorbox}
\DeclareUnicodeCharacter{2606}{$\star$}

\AtBeginDocument{%
  \providecommand\BibTeX{{%
    Bib\TeX}}}
    
\begin{document}

\newcommand{\attackname}{{MIND-Crypt}}

\newcommand{\jimmy}[1]{%
  \textcolor{red}{\textbf{Jimmy's comment:}} \textcolor{red}{#1}%
}

\title{A Machine Learning-Based Framework for Assessing Cryptographic Indistinguishability of Lightweight Block Ciphers}


\author{
\IEEEauthorblockN{Jimmy Dani, Kalyan Nakka, Nitesh Saxena}
\IEEEauthorblockA{
Department of Computer Science, Texas A\&M University \\
College Station, TX, USA \\
\{danijy, kalyan, nsaxena\}@tamu.edu
}
}


\pagestyle{plain}

\maketitle

\begin{abstract}
Indistinguishability is a fundamental principle of cryptographic security, crucial for securing data transmitted between Internet of Things (IoT) devices.
This principle ensures that an attacker cannot distinguish between the encrypted data, also known as ciphertext, and random data or the ciphertexts of the two messages encrypted with the same key. This research investigates the ability of machine learning (ML) in assessing indistinguishability property in encryption systems, with a focus on lightweight ciphers. As our first case study, we consider the SPECK32/64 and SIMON32/64 lightweight block ciphers, designed for IoT devices operating under significant energy constraints.

In this research, we introduce \attackname\footnote{We refer to our attack framework as \attackname\, which stands for ``\textbf{\textit{M}}achine learning based framework for assessing  \textbf{\textit{IND}}istinguishability of \textbf{\textit{Crypt}}ographic Algorithms.''}, a novel ML-based framework designed to assess the cryptographic indistinguishability of lightweight block ciphers, specifically the SPECK32/64 and SIMON32/64 encryption algorithm in CBC mode (Cipher Block Chaining), under Known Plaintext Attacks (KPA). Our approach involves training ML models using ciphertexts from two plaintext messages encrypted with same key to determine whether ML algorithms can identify meaningful cryptographic patterns or leakage. Our experiments show that modern ML techniques consistently achieve accuracy equivalent to random guessing, indicating that no statistically exploitable patterns exists in the ciphertexts generated by considered lightweight block ciphers. Furthermore, we demonstrate that in ML algorithms with all the possible combinations of the ciphertexts for given plaintext messages reflects memorization rather than generalization to unseen ciphertexts.

Collectively, these findings suggest that existing block ciphers have secure cryptographic designs against ML-based indistinguishability assessments, reinforcing their security even under round-reduced conditions.

\end{abstract}


\section{Introduction}
\label{lab:intro}

Indistinguishability is the basis for building secure encryption systems. Concretely, indistinguishability means that the adversary can not tell the difference between the ciphertexts corresponding to two plaintexts with a probability significantly better than 0.50. It is an important notion underlying encryption security since it implies that the adversaries are unable to decipher any useful information about the plaintext given the ciphertext. Moreover, a broken indistinguishability property exposes deterministic or predictable patterns in the encryption process, making the system susceptible to more effective attacks, such as ciphertext-only attacks where the plaintext is deciphered without the key. This not only undermines the trust and reliability of the cryptographic system but also paves the way for practical decryption techniques that could exploit this predictability. Therefore, preserving indistinguishability is essential to maintain the overall integrity and security of encryption schemes.

\noindent \textbf{Lightweight Block Ciphers.} The Internet of Things (IoT) exemplifies a domain where cryptography's vital role is particularly pronounced, due to its explosive growth and the evolving capabilities of connected devices. With projections estimating about 40 billion devices connected by 2030 \cite{iot-stats, cisco-intro-blog-1, statista-iot-stats, iot-stats-2}, the diversity of applications—from smart home devices enhancing residential convenience and security, to advanced systems in healthcare monitoring and industrial IoT (IIoT)—is transforming traditional industries. However, many IoT devices operate under constraints of processing power and memory, necessitating cryptographic solutions that optimize security without imposing significant computational burdens. Among lightweight block ciphers, the SPECK32/64 and SIMON32/64 ciphers, designed by the National Security Agency, stands out for its operational efficiency and simplicity, tailored specifically to meet the needs of these resource-constrained environments \cite{techcrunchInternetThings, AMBA16, beaulieu2015simon}.

\noindent \textbf{Cryptanalysis and Machine Learning.} As cryptographic systems evolve in complexity and sophistication, so too does cryptanalysis -- the study and practice of deciphering codes, ciphers, and encrypted messages without the use of actual key. This discipline has seen significant advancements through a variety of techniques, reflecting the ongoing arms race between cryptography and cryptanalysis. Traditional methods such as side-channel attacks \cite{ZY-05-sca, RM20-sca, KJ98-sca, PRC-06-sca}, fault injection attacks \cite{DJM-11-FI, CJA-02-FI, BA-12-FI, SC-23-FI}, mathematical analysis \cite{AS21-cryptanalysis-classical-ciphers, LSNS-22-cryptanalysis-ecc, XY-19-cryptanalysis-neural}, and brute-force attacks \cite{GA22-cryptanalysis-brute-force, SSKM-22, VR22, MCJ-19-bruteforce} have continually been refined in tandem with advancements in cryptographic techniques. However, as cryptographic algorithms become more complex, the effectiveness of these traditional approaches is increasingly challenged, necessitating newer methodologies. This evolving landscape has sparked considerable interest in integrating machine learning with cryptanalysis, offering novel approaches to breaking cryptographic systems and presenting new challenges to their robustness.

In 2019, Gohr \cite{GA19} proposed a differential attack on round-reduced SPECK32/64, focusing on the development of neural distinguishers that could effectively distinguish ciphertexts differing by a specific difference delta from random text. This approach leveraged DL, specifically deep residual neural networks, which demonstrated superior performance compared to traditional cryptographic distinguishers. Further enhancing the practicality of his method, Gohr integrated a novel key search policy based on Bayesian optimization, significantly improving the efficiency of key recovery processes. Following Gohr's work, Benamira et al. \cite{ABDGTHQQT21-gohr-follow-up} conducted detailed analysis and showed neural distinguisher developed by Gohr generally relies on the differential distribution on the ciphertext pairs, but also on the differential distribution in penultimate and antepenultimate rounds. 
This approach not only showcased DL's potential in enhancing traditional cryptanalysis but also emphasizes the need to probe deeper into the cipher's behavior by exploring the notion of indistinguishability. Unlike prior research focused primarily on differential cryptanalysis, our approach uniquely targets indistinguishability—an essential property underpinning robust encryption—and systematically assess it against advanced machine learning methods.

\begin{tcolorbox}[colframe=black!50!black, colbacktitle=black!40!white, 
coltitle=black, width=\columnwidth]
    Our research investigates the potential of ML techniques to assess the
indistinguishability of lightweight block ciphers, specifically SPECK32/64 and SIMON32/64. Compromising indistinguishability renders the cipher fundamentally insecure. This process involves training a deep learning model on ciphertexts from two distinct messages, $\mathcal{P}_{1}$ and $\mathcal{P}_{2}$, and aims to determine if a challenge ciphertext belongs to message $\mathcal{P}_{1}$ or $\mathcal{P}_{2}$. 
\end{tcolorbox}

\noindent \textbf{Focus of Our Research.} In contrast to Gohr \cite{GA19}, our research shifts the focus from differential attack strategies to the broader concept of indistinguishability within lightweight block ciphers (e.g., SPECK32/64, and SIMON32/64). Unlike Gohr’s approach, which targets specific, known differential paths for key recovery, our study employs ML to assess whether a model can distinguish between ciphertexts of two plaintext messages encrypted using the same key. Our analysis demonstrates that achieving a generalized ML-based indistinguishability is fundamentally more challenging than exploiting predefined differential characteristics. Consequently, our results highlight that existing lightweight block ciphers remain robust, as current ML methods fail to compromise their indistinguishability.

To illustrate the practical implications of our research, consider a scenario involving a smart home security system that utilizes the SPECK32/64 or SIMON32/64 cipher to encrypt data from sensors such as motion detectors and window sensors. If indistinguishability were compromised, an adversary might differentiate encrypted sensor signals, distinguishing, for instance, whether ciphertext originates from motion sensors detecting indoor movement or window sensors detecting window openings. Such an ability would pose severe privacy risks, enabling unauthorized parties to infer sensitive patterns (e.g., movements), without explicitly decrypting the messages.

Formally, in our study, we address the following research question: \textit{Can ML techniques compromise the indistinguishability property of lightweight block ciphers?} Our findings provide strong evidence that current lightweight block cipher implementations are secure against ML-based indistinguishability assessments.

When designing \attackname, we considered assumptions typical of the Known Plaintext Attack (KPA) scenario, where the attacker has access to both plaintexts and their corresponding ciphertexts encrypted under the same key. Here, the primary focus of an attacker is to identify if the challenge ciphertext belongs to message $\mathcal{P}_{1}$ or $\mathcal{P}_{2}$, thus testing the fundamental indistinguishability of the considered encryption schemes. Our objective is not to demonstrate vulnerability but to investigate whether subtle leakages might be exploited by ML. We study both its standard configuration and round-reduced versions to understand if these variations affect resistance to ML.

\noindent \textbf{Our Methodology \& Experiments.} We approach this challenge by framing the task as a binary classification problem, where the ML classifier is trained on  previously-known ciphertexts $\mathcal{C}_{1}$ and $\mathcal{C}_{2}$ corresponding to two fixed plaintexts $\mathcal{P}_{1}$ and $\mathcal{P}_{2}$, respectively, and using the trained model to predict whether any new challenge ciphertexts correspond to $\mathcal{P}_{1}$ or $\mathcal{P}_{2}$. To train the model, the attacker generates ciphertexts of these messages by encrypting them under the same key.

Our experiments show that the performance of the ML models remains consistently around random guessing levels ($\approx$50\%). These findings suggests that ML models are unable to extract meaningful patterns from ciphertexts produced by lightweight encryption schemes. Consequently, our results emphasize that ML-techniques, despite their advanced capabilities, cannot challenge the indistinguishability property cryptographic algorithms.

\noindent\textbf{Our Contributions and Summary of Results:} The main contributions and findings are summarized as follows:
\begin{enumerate} [leftmargin=0.5cm]
    \item \textbf{\textit{A Novel Machine Learning Framework:}} We designed \attackname, a novel machine learning-based framework that utilized ML techniques to investigate the indistinguishability of lightweight block ciphers. More specifically, we leverage DL to implement \attackname.

    \item \textbf{\textit{Comprehensive Evaluation of Cryptographic Indistinguishability:}} We evaluate cryptographic indistinguishability of popular lightweight block ciphers by leveraging multiple state-of-the-art deep learning architectures, including ResNet, CNN, LSTM, and BiLSTM. Our experiments demonstrate that all evaluated ML models consistently achieve accuracies equivalent to random guessing ($\approx$50\%), clearly indicating their inability to detect meaningful cryptographic leakage or statistical patterns. 

    \item \textbf{\textit{Analysis of Memorization vs. Generalization:}} We provide a detailed analysis distinguishing memorization from generalization in DL model predictions, leveraging reduced-entropy datasets specifically designed to study memorization effects.

    \item \textbf{\textit{Security Assurance for IoT Devices:}} Our results provides practical assurance, demonstrating that lightweight block ciphers such as SPECK32/64 and SIMON32/64 are secure  against ML-based indistinguishability attacks in realistic, resource-constrained IoT environments.
\end{enumerate}

\noindent
\textbf{Reproducibility.} we will make code and datasets publicly available upon the publication of this research.
\section{Background \& Preliminaries}
\label{sec:preliminaries}

\subsection{Lightweight Block Ciphers}
\subsubsection{SPECK32/64 Block Cipher}
SPECK is a family of lightweight block ciphers, denoted as SPECK$M/N$ where \textit{M}, \textit{N} are block size and key size respectively in bits, developed by Beaulieu, Treatman-Clark, Shors, Weeks, Smith and Wingers \cite{beaulieu2015simon2} for NSA. It is an add-rotate-xor (ARX) cipher with operations like modular addition (mod $2^k$) $\boxplus$, bitwise addition $\oplus$, and bitwise rotation (left $\ll$ and right $\gg$) applied on \textit{k}-bit words, aimed to build efficient cipher for software implementations in IoT devices \cite{beaulieu2015simon}. The round function of SPECK $F : \mathbb{F}_2^{2k} \times \mathbb{F}_2^{2k} \rightarrow \mathbb{F}_2^{2k}$, computes the next round state $(L_{i+1}, R_{i+1})$ using a \textit{k}-bit subkey \textit{K} and current round state $(L_i, R_i)$ as, $L_{i+1} = ((L_i \gg \alpha) \boxplus R_i) \oplus K$, and $R_{i+1} = (R_i \ll \beta) \oplus L_{i+1}$.

Here, $\alpha, \beta$ are rotation constants ($\alpha = 7, \beta = 2$ for SPECK32/64 and $\alpha = 8, \beta = 3$ for remaining). The ciphertext is produced from the input plaintext by employing this round function for a fixed number of times (22 rounds for SPECK32/64). Further, the design of SPECK32/64 balances security with minimal computation overhead making it an ideal candidate for studying indistinguishability in resource constrained IoT devices \cite{AMBA16, beaulieu2015simon}.

\subsubsection{SIMON32/64 Block Cipher}
SIMON is a family of lightweight block ciphers, denoted as SIMON$M/N$, where \textit{M} represents the block size in bits, and \textit{N} denotes the key size in bits. SIMON was designed by Beaulieu, Shors, Smith, Treatman-Clark, Weeks, and Wingers for the NSA \cite{beaulieu2015simon2}, specifically optimized for efficient implementation in hardware-constrained environments, such as embedded systems \cite{beaulieu2015simon}. SIMON employs a balanced Feistel network structure, particularly suited for hardware efficiency due to its simplicity, minimal gate count, and compact area utilization.

For SIMON32/64, the cipher employs a word size of 16 bits (thus a 32-bit block size) and a 64-bit key. The SIMON32/64 variant uses 32 rounds of encryption, providing adequate security for resource-constrained devices. The minimalistic and serialized design makes it highly suitable for hardware implementations where area minimization and power efficiency are critical, such as embedded IoT platforms \cite{beaulieu2015simon, AMBA16}.

\section{Threat Model \& Assumptions}
\label{lab:threat-model}


Our study investigates the security of the SPECK32/64 and SIMON32/64 lightweight block ciphers in CBC mode (Cipher Block Chaining) against Known Plaintext Attacks (KPA). We primarily focus on an attacker's ability to distinguish between the ciphertexts of two different messages encrypted using the same key. This is particularly relevant for IoT devices that operate under significant energy constraints and require efficient and lightweight cryptographic solutions like the SPECK32/64 or SIMON32/64 cipher.

In our attack model, we consider a passive attack scenario where the attacker gains excess ciphertexts, all encrypted with the same key, without performing active attacks such as Chosen-Ciphertext Attacks (CCA). To illustrate the practical implications of violating indistinguishability (briefly noted in Section \ref{lab:intro}) in cryptographic systems, consider a smart home security system that uses the SPECK32/64 or SIMON32/64 lightweight block cipher to encrypt data from various constrained IoT sensors around the house. These sensors -- including motion detectors, cameras, and window sensors -- regularly send encrypted data to a central monitoring system. Adopting a passive attack scenario enhances the practical relevance of our assessment, as it represents a realistic threat where attackers merely observe ciphertexts without active manipulations, commonly encountered in practical IoT security environments.

Mathematically, we denote the plaintext by $\mathcal{P}$, the ciphertext by $\mathcal{C}$, and the secret key by $\mathcal{K}$. The encryption function $\mathcal{E}_{K}$ uses the key $\mathcal{K}$ to transform plaintext into ciphertext. A cipher maintains indistinguishability if no polynomial-time adversary can distinguish between the ciphertexts of two different plaintexts encrypted with the same key with a probability significantly better than 0.5. 

The attacker selects two different fixed plaintexts, $\mathcal{P}_{1}$ and $\mathcal{P}_{2}$ (e.g., ``heat'' or ``cool'' commands that adjusts the temperature using thermostat), which are encrypted using the same secret key $\mathcal{K}$, resulting in ciphertexts $\mathcal{C}_{1}$ and $\mathcal{C}_{2}$. Subsequently, the attacker employs a deep learning model, trained with multiple instances of ciphertexts $\mathcal{C}_{1}$ and $\mathcal{C}_{2}$. This model is then utilized to classify new challenge ciphertexts, determining whether they correspond to $\mathcal{P}_{1}$ or $\mathcal{P}_{2}$, potentially breaching the indistinguishability property of the encryption scheme.

Our model extends these concepts by allowing the attacker to simulate data generation without direct access, avoiding the active manipulation typical of CCA. The attacker aims to identify patterns, anomalies, or relationships in the ciphertexts that differentiate those corresponding to two distinct, same-byte-length plaintexts. Successfully differentiating ciphertexts beyond chance agreement signifies vulnerabilities in the block cipher, whereas failure to do so would validate the cipher's robustness under passive attack settings.

\section{\attackname: Design \& Methodology}
\label{sec:methodology}
In this section, we introduce \attackname, a machine learning-based assessment framework designed to evaluate the cryptographic indistinguishability of lightweight block ciphers, specifically SPECK32/64 and SIMON32/64, operating in Cipher Block Chaining (CBC) mode. 

\subsection{Framework Design}
Our primary objective is to investigate whether machine learning (ML) algorithms can identify statistically meaningful patterns or cryptographic leakage in ciphertexts generated by these lightweight block ciphers. Deep learning models have demonstrated significant promise for solving complex classification problems in cybersecurity, such as malware detection, intrusion detection, and traffic classification. In this study, we utilized multiple DL architectures, namely, Convolution Neural Networks~\cite{LeCun-CNN} (CNNs), Long-Short Term Memory (LSTM)~\cite{LSTM-1} networks, Bidirectional LSTM (BiLSTM)~\cite{BILSTM-Graves-05} networks, and Residual Neural Networks (ResNets)~\cite{GA19} to comprehensively evaluate cryptographic indistinguishability of lightweight block ciphers. The details of these DL architectures is as follows:

\subsubsection{Convolutional Neural Networks (CNNs)} CNNs, introduced by LeCun et al. \cite{LeCun-CNN} can effectively extract hierarchical spatial features from input data via convolutional layers. CNNs leverage multiple convolutional layers to automatically identify hierarchical patterns within the input data, which reduces the reliance on manual feature extraction. Although CNNs have historically been applied extensively in image recognition tasks, their capability to capture subtle local statistical dependencies also makes them well-suited for security research. CNNs are highly effective for classification tasks involving structures, grid-like data. These models have successfully improved classification accuracy for security problems such as network intrusion detection \cite{li2020-ids-1} and malware analysis \cite{Nethala-CNN}. 

\subsubsection{Long-Short Term Memory (LSTM)} LSTMs were introduced by Hochreiter and Schmidhuber \cite{HSSJ-LSTM-97} are a type of recurrent neural network capable of learning sequential dependencies and long-term temporal patterns. LSTM architectures employ specialized gating mechanisms that include input, output and forget gates to effectively preserve long-range dependencies within sequential data, addressing the vanishing gradient problem common to traditional RNNs. For ciphertext indistinguishability assessment, the sequential characteristics of ciphertext bits are critically important. The intrinsic ability of LSTM networks to capture long-range sequential patterns makes them particularly suitable for analyzing cryptographic ciphertexts generated through block ciphers.

\subsubsection{Bidirectional Long-Short Term Memory (BiLSTM)} BiLSTM network architecture proposed by Graves and Schmidhuber \cite{BILSTM-Graves-05} enhancs traditional LSTM architectures by simultaneously processing input sequences in both forward and backward directions. This bidirectional processing allows BiLSTM networks to leverage past and future context at each point in a given sequence, significantly improving their ability to capture complex dependencies. In cryptographic indistinguishability analysis, the direction-agnostic nature of BiLSTMs may offer additional sensitivity in detecting subtle statistical differences across ciphertext sequences, thereby providing a comprehensive evaluation capability for the presence or absence of cryptographic leakage or patterns.

\subsubsection{Residual Neural Networks (ResNets)}
He et al. \cite{HKZX-16-resnet} introduced ResNets to address the vanishing gradients problem in deep neural network (DNN) training by utilizing residual blocks. These blocks, featuring stacked convolutional layers with skip connections, allow the network to learn residual functions, focusing on differences rather than complete transformations. ResNets have been successfully applied in various security applications \cite{SCKGKK-24-resnet, AAVP-22-resnet, YSRQ-22-resnet, CSVVSI-16-resnet}.

In cryptanalysis, ResNet models are effective at identifying complex patterns, which helps with tasks such as automated cipher breaking and differential cryptanalysis. Their architecture allows for more accurate and efficient prediction of differential characteristics, enhancing encryption analysis and vulnerability insights. A prominent example is the work of Gohr et al. \cite{GA19}, who leveraged deep residual neural networks to identify differential characteristics in round-reduced versions of lightweight block ciphers such as SPECK32/64. Their findings highlighted that ResNets could surpass traditional cryptanalytic methods in specific scenarios involving reduced cipher complexity. Adrien et al. \cite{ABDGTHQQT21-gohr-follow-up} discuss how machine learning, including ResNets, advances cryptanalytic and cyber defense techniques.

\begin{figure*}[htp!]
  \centering
  \includegraphics[scale=0.5]
  {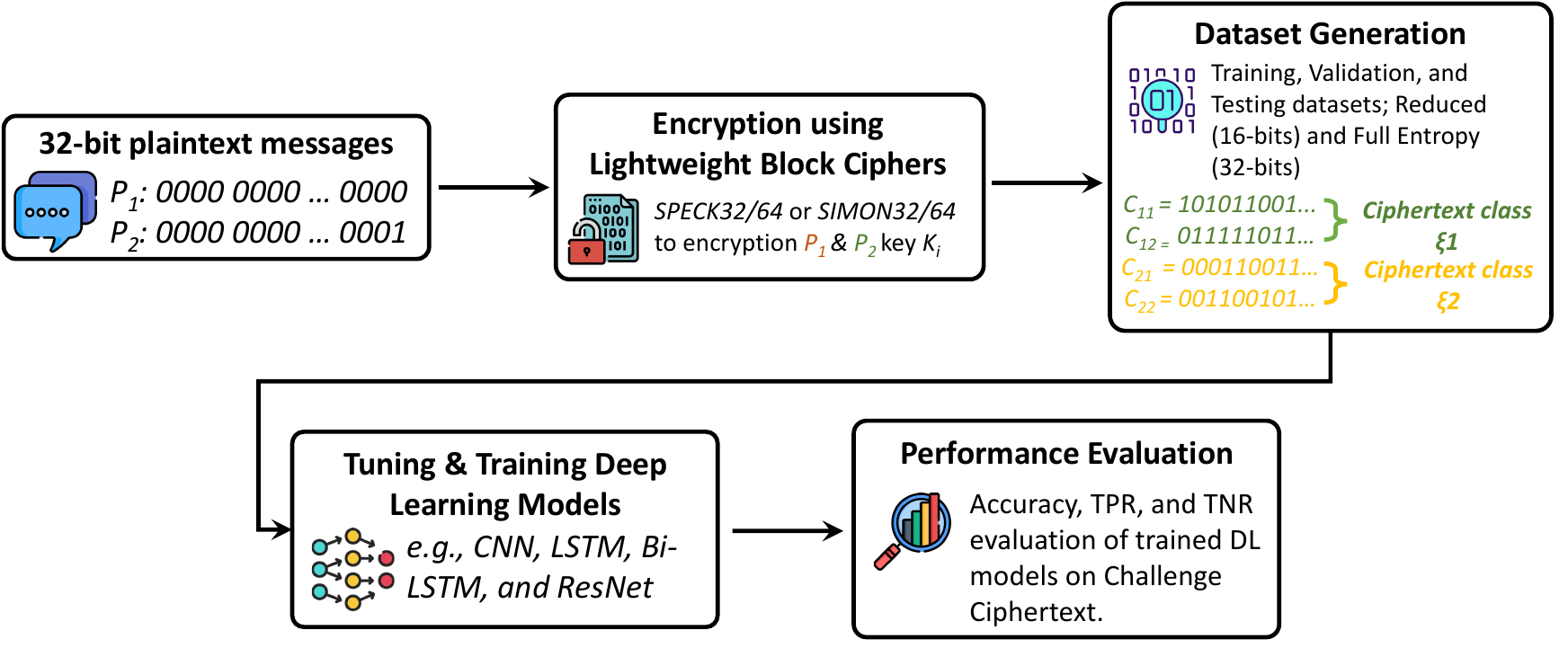}
  \vspace*{-1mm}
  \caption{The \attackname\ assessment framework - Investigating the indistinguishability of SPECK32/64 and SIMON32/64 lightweight block ciphers. Two plaintext messages encrypted with the same key, represented in binary format, form the basis for training a DL model.}
  \label{fig:system-design}
\end{figure*}

\subsection{Framework Implementation}
Figure~\ref{fig:system-design} illustrates our assessment framework, detailing the entire process from message selection and ciphertext generation to ML-based assessment. Initially, two plaintext messages $\mathcal{P}_{1}$ and $\mathcal{P}_{2}$, each having byte-length and differing by exactly one bit, are encrypted multiple times using either SPECK32/64 or SIMON32/64 ciphers under a fixed encryption key \textit{k} with CBC mode. Our DL models are trained for binary classification task of separating ciphertexts into two classes: $\xi_{1}$ and $\xi_{2}$. To explain, $\xi_{1}$ includes the ciphertexts of $\mathcal{P}_{1}$, labeled as $\mathcal{C}{1}_{i}$ ($\mathcal{C}{1}_{i} = Enc_{k} (\mathcal{P}_{1}$)), where $i \in \{1, 2, \ldots, n\}$. Similarly, $\xi_{2}$ includes the ciphertexts of $\mathcal{P}_{2}$, labeled as $\mathcal{C}{2}_{i}$ ($\mathcal{C}{2}_{i} = Enc_{k} (\mathcal{P}_{2}$)) for $i \in \{1, 2, \ldots, n\}$. It should be noted that the Initialization Vectors (IVs) are used only as a part of encryption process, and not included in the training data for the DL model. This design choice ensures that the model learns to identify any intrinsic properties or subtle differences in the ciphertext generated from $\mathcal{P}_{1}$ and $\mathcal{P}_{2}$, without relying on external factor of the IVs.

Following ciphertext generation, we convert the ciphertexts into binary format, adhering to the data preparation methods described by Gohr et al. \cite{GA19} for examining differential attacks on SPECK32/64. Utilizing this methodology, we feed these binary ciphertexts into a DL model. While Gohr et al.~\cite{GA19} demonstrated the effectiveness of ResNet models in identifying differential characteristics within ciphertexts, their approach primarily leveraged spatial hierarchical features through convolutional residual blocks. To thoroughly assess cryptographic indistinguishability, we employ diverse DL architectures capable of capturing different types of patterns or subtle biases within ciphertext data. Specifically, we selected CNN architectures for their proven efficiency in extracting spatial and local feature patterns. Additionally, we included LSTM and BiLSTM networks due to their capability to detect sequential dependencies and temporal correlations that might remain undetected by purely convolution-based architectures. The combination of spatial (CNN), sequential (LSTM/BiLSTM), and hierarchical (ResNet) learning mechanisms ensures a robust, multi-dimensional analysis, providing comprehensive insights into security of lightweight block ciphers against varied ML-based cryptanalytic approaches. Each DL model in our framework is trained for binary classification to distinguish ciphertexts derived from plaintexts $\mathcal{P}_{1}$ and $\mathcal{P}_{2}$.

Finally, the trained ML models are evaluated on unseen challenge ciphertext samples. By analyzing model predictions and systematically comparing their performance against a random guessing baseline ($\approx$50\% accuracy), we provide empirical insights into whether state-of-the-art ML techniques can uncover meaningful cryptographic vulnerabilities. Rather than demonstrating exploitable weakness, our comprehensive assessment highlight the robustness of lightweight block cipher designs against ML-based indistinguishability attacks.

\section{Assessing Lightweight Block Ciphers using \attackname}
\label{lab:evaluation}
In this section, we describe how our proposed \attackname\ framework can be utilized for assessing lightweight block ciphers. We describe the datasets, experiment settings, and evaluation metric considered for assessing our framework.

\subsection{Description of the Dataset}
In our study, we evaluated the effectiveness of the \attackname\ by utilizing a publicly available implementation of SPECK32/64 provided by Gohr \cite{GA19}, and SIMON32/64 implementation \cite{simon-implementation}. Our objective was to investigate the principle of indistinguishability, which required control over the inputs provided for encryption. In our experimental setup, we aimed to align closely with the methodologies previously established by Gohr, particularly regarding the generation of cryptographic components. 

Additionally, to distinguish between memorization and generalization behaviors exhibited by ML models, we conducted a proof-of-concept evaluation using a simplified cryptographic setup. Specifically, we intentionally reduced the entropy in the SPECK32/64 encryption algorithm from the standard 32 bits to 16 bits. This reduction created an artificially weakened cryptographic scenario, significantly decreasing the complexity and thereby increasing the potential for identifiable statistical patterns. We emphasize that this simplified experiment was conducted solely for analyzing ML model behaviors regarding memorization versus genuine generalization, and was not intended as a realistic assessment of the cipher's actual indistinguishability or security under standard cryptographic conditions.

To this end, we made several modifications to the original SPECK32/64 implementation provided in \cite{GA19}:

\begin{enumerate} [leftmargin=0.5cm, topsep=2pt]
    \item \textit{\textbf{Encryption Mode:}} We shifted from the Electronic Code Book (ECB) mode used in Gohr's original code to CBC mode. This change involved encrypting the messages using CBC with randomly generated initialization vectors (IVs) and applying an XOR operation to the messages before encryption.

    \item \textit{\textbf{Key Usage:}} Unlike the original implementation that used varying keys, we utilized a single, fixed key securely generated using Gohr’s methodology. This consistency was vital for comparing the indistinguishability of outputs. This approach allowed us to isolate the impact of message variation on ciphertext indistinguishability without key variability influencing the results.

    \item \textit{\textbf{Generating IVs:}} We employed the \texttt{frombuffer} module in \texttt{NumPy} library in conjunction with Python's \texttt{os.urandom} to generate cryptographically secure IVs, mirroring Gohr's method.

    \item \textit{\textbf{Correctness:}} To ensure the correctness of our modifications, we decrypted the ciphertexts to verify that they reverted accurately to the original plaintexts, labeled `0' and `1'.

    \item \textit{\textbf{Message Selection:}} We chose two specific messages of identical 32-bit length, differing by only a single bit at the binary level, labeled `0' and `1'. This allowed us to directly assess the effect of minimal input variation on the encryption output.

\end{enumerate}

For exploring indistinguishability using DL, we collected $10^{7}$ training samples, $10^{6}$ samples each for validation and testing across `$\mathcal{R}$' rounds of encryption schemes. Each dataset segment maintained an equal number of samples from two classes, representing ciphertexts of two distinct plaintext messages encrypted with the same key. The training data was used to train a DL model, while the testing data was utilized to evaluate the performance of the trained model on an unseen dataset. This allowed the DL model to detect subtle differences in ciphertexts of the selected messages. To facilitate the learning process, the ciphertexts were represented as 32-bit binary vectors, providing a consistent input format for the DL.

\subsection{Experiment Settings}
The implementation of the \attackname\ was conducted using the Python programming language, leveraging the open-source library TensorFlow \cite{tensorflow2015-whitepaper} for the development, training, and evaluation of the deep learning model. To optimize the neural network's hyperparameters, we employed Optuna \cite{optuna_2019}, a software framework designed for efficient and automatic hyperparameter optimization. Specifically, we utilized Optuna's \texttt{TPESampler}, which implements the Tree-structured Parzen Estimator (TPE) algorithm—a Bayesian optimization approach that models the objective function using two separate densities to efficiently navigate the hyperparameter search space \cite{tpe-search}. The hyperparameter search process was configured to execute up to 100 trials or terminate if the search duration exceeded 200 hours. The search space for the hyperparameters is detailed in Table~\ref{tab:model_specific_hyperparams} and ~\ref{tab:common_hyperparams}.

\begin{table*}[htbp]
\centering
\caption{Model-Specific Hyperparameter Search Space}
\begin{tabularx}{\textwidth}{lXX}
\toprule
\textbf{Hyperparameter} & \textbf{LSTM-based Models (LSTM, BiLSTM)} & \textbf{CNN-based Model (1D CNN)} \\
\midrule
\texttt{No. of LSTM Layers} & \{2, 3, 4, 5, 6, 7, 8, 9\} & -- \\
\texttt{LSTM Cells in Each Layer}     & \{200, 300, 400, 500\} & -- \\
\texttt{No. of Convolution Layers} & -- & \{2, 3, 4, 5, 6, 7, 8, 9\} \\
\texttt{No. of Filters}    & -- & \{2, 4, 8, 16, 32, 64, 128, 256\} \\
\texttt{Kernel Size}    & -- & \{2, 3, 4, 5, 7, 9, 11, 13, 15, 17, 19, 21\} \\
\texttt{Convolution Stride Size} & -- & \{2, 3, 4, 5, 7, 9, 11, 13, 15, 17, 19, 21\} \\
\texttt{Pool Size}      & -- & \{1, 2, 3, 4\} \\
\texttt{Pool Stride Size} & -- & \{2, 3, 4\} \\
\bottomrule
\end{tabularx}
\label{tab:model_specific_hyperparams}
\vspace{-2mm}
\end{table*}


\begin{table}[htbp]
\centering
\caption{Common Hyperparameter Search Space Across All Models}
\begin{tabularx}{\linewidth}{lX}
\toprule
\textbf{Hyperparameter} & \textbf{Search Space} \\
\midrule
\texttt{decay} & \{0.05, 0.1, 0.2, 0.3\} \\
\texttt{Dropout Rate} & \{0.05, 0.1, 0.2, 0.3, 0.4\} \\
\texttt{Activation Function} & \{Softsign, ELU, Selu, ReLU, Tanh\} \\
\texttt{No. of Dense Layers} & \{1, 2, 3, 4, 5, 6, 7, 8, 9\} \\
\texttt{No. of Neurons in FC Layer} & \{256, 512, 1024, 2048, 4096\} \\
\texttt{Activation Function in FC Layer} & \{Softsign, ELU, Selu, ReLU, Tanh\} \\
\texttt{Dropout Rate FC} & \{0.05, 0.1, 0.2, 0.3, 0.4\} \\
\texttt{optimizer} & \{RMSprop, Adagrad, Adam, Adamax, Nadam, SGD\} \\
\texttt{Epochs} & \{100, 200, 300\} \\
\texttt{Batch Size} & \{256, 512, 1024\} \\
\texttt{Learning Rate} & [0.00001, 0.01] (log scale) \\
\bottomrule
\end{tabularx}
\label{tab:common_hyperparams}
\end{table}

\noindent
\textbf{DL Model Training for Indistinguishability Assessment.} To study cryptographic indistinguishability of ciphertexts, we implemented and trained four distinct DL architectures: ResNets \cite{GA19}, CNN, LSTM, and BiLSTM networks. The ResNet architecture developed by Gohr was specifically selected due to its success in identifying differential characteristics in reduced-round versions of SPECK32/64 cipher. We adapted Gohr's ResNet model for our binary classification task. This adaptation aimed to assess whether machine learning could effectively distinguish ciphertexts generated from two distinct plaintexts, $\mathcal{P}_{1}$ and $\mathcal{P}_{2}$ encrypted using same key.

We extended the assessment to include CNN, LSTM, and BiLSTM architectures, commonly employed in image and sequence processing. These models were adapted to process ciphertext data by converting inputs into binary vector representations, facilitating sequential (LSTM/BiLSTM) or spatial (CNN) feature extraction. This methodology ensured a comparative analysis of DL architectures in the context of ciphertext indistinguishability. Detailed architectural specifications and hyperparameter settings for the ResNet model are available in Gohr \cite{GA19}. This work examines the potential of DL techniques to serve as distinguishers, contributing to the broader understanding of cryptographic security in the using machine learning.

\subsection{Evaluation Metrics}
To evaluate the efficacy of the \attackname\ across different settings, we performed a comprehensive assessment using a DL model to classify ciphertexts into two distinct classes, $\xi_{1}$ and $\xi_{2}$. This evaluation employs three key metrics: Accuracy, True Positive Rate (TPR), and True Negative Rate (TNR), similar to the metrics considered in studies by \cite{GA19, ABDGTHQQT21-gohr-follow-up} that explore differential attacks in the SPECK32/64 encryption scheme. Furthermore, accuracy, TPR, and TNR were specifically chosen because they collectively provide clear insights into model biases, detection capabilities, and overall effectiveness in distinguishing ciphertext classes.

Accuracy gauges the model's overall effectiveness at correctly classifying ciphertexts belonging to class $\xi_{1}$ or $\xi_{2}$. We calculate accuracy as the proportion of correct classification—both true positives and true negatives—out of the total ciphertexts examined. A higher accuracy value reflects superior model performance in discriminating accurately between ciphertexts associated with classes $\xi_{1}$ and $\xi_{2}$.

True Positive Rate (TPR), or sensitivity, specifically measures the model's precision in identifying ciphertexts that genuinely belong to class $\xi_{1}$. This metric is crucial for cryptographic applications as it reflects the model's ability to capture the unique characteristics expected from $\xi_{1}$ under particular encryption conditions. High TPR is vital, especially in situations where failing to correctly identify a ciphertext from $\xi_{1}$ could pose significant security threats.

True Negative Rate (TNR), or specificity, evaluates the model's accuracy in classifying ciphertexts into class $\xi_{2}$ when they do not belong to class $\xi_{1}$. This measure is essential for ensuring the model effectively identifies ciphertexts that do not adhere to the characteristics of class $\xi_{1}$, thus preventing false positives. A high TNR underscores the model's reliability in excluding non-conforming encryption outputs, pivotal for upholding robust cryptographic defenses.

In addition to these key metrics, we also provided detailed analysis using Precision, Recall, F1-Score, Receiver Operating Characteristic Area Under the Curve (ROC-AUC), False Negative Rate (FNR), and False Positive Rate (FPR).

\begin{table*}[htp!]
\centering
\small
\caption{Indistinguishability assessment for SPECK32/64 and SIMON32/64 in Round-Reduced Standard Configuration using \attackname}
\label{tab:dl-results}
\begin{tabular}{|ccccccccccc|}
\hline
\multicolumn{11}{|c|}{\textbf{Round Reduced}} \\ \hline
\multicolumn{1}{|c|}{\textbf{Cipher}} &
  \multicolumn{1}{c|}{\textbf{DL Model}} &
  \multicolumn{1}{c|}{\textbf{Accuracy}} &
  \multicolumn{1}{c|}{\textbf{Precision}} &
  \multicolumn{1}{c|}{\textbf{Recall}} &
  \multicolumn{1}{c|}{\textbf{F1-Score}} &
  \multicolumn{1}{c|}{\textbf{ROC-AUC}} &
  \multicolumn{1}{c|}{\textbf{TPR}} &
  \multicolumn{1}{c|}{\textbf{TNR}} &
  \multicolumn{1}{c|}{\textbf{FPR}} &
  \textbf{FNR} \\ \hline
\multicolumn{1}{|c|}{\multirow{4}{*}{\textbf{SPECK32/64}}} &
  \multicolumn{1}{c|}{ResNet} &
  \multicolumn{1}{c|}{0.5000} &
  \multicolumn{1}{c|}{0.0000} &
  \multicolumn{1}{c|}{0.0000} &
  \multicolumn{1}{c|}{0.0000} &
  \multicolumn{1}{c|}{0.5008} &
  \multicolumn{1}{c|}{0.0000} &
  \multicolumn{1}{c|}{1.0000} &
  \multicolumn{1}{c|}{0.0000} &
  1.0000 \\ \cline{2-11} 
\multicolumn{1}{|c|}{} &
  \multicolumn{1}{c|}{CNN} &
  \multicolumn{1}{c|}{0.5003} &
  \multicolumn{1}{c|}{0.5043} &
  \multicolumn{1}{c|}{0.0356} &
  \multicolumn{1}{c|}{0.0665} &
  \multicolumn{1}{c|}{0.5005} &
  \multicolumn{1}{c|}{0.0355} &
  \multicolumn{1}{c|}{0.9650} &
  \multicolumn{1}{c|}{0.0350} &
  0.9644 \\ \cline{2-11} 
\multicolumn{1}{|c|}{} &
  \multicolumn{1}{c|}{LSTM} &
  \multicolumn{1}{c|}{0.5000} &
  \multicolumn{1}{c|}{0.0000} &
  \multicolumn{1}{c|}{0.0000} &
  \multicolumn{1}{c|}{0.0000} &
  \multicolumn{1}{c|}{0.5014} &
  \multicolumn{1}{c|}{0.0000} &
  \multicolumn{1}{c|}{1.0000} &
  \multicolumn{1}{c|}{0.0000} &
  1.0000 \\ \cline{2-11} 
\multicolumn{1}{|c|}{} &
  \multicolumn{1}{c|}{BiLSTM} &
  \multicolumn{1}{c|}{0.5000} &
  \multicolumn{1}{c|}{0.0000} &
  \multicolumn{1}{c|}{0.0000} &
  \multicolumn{1}{c|}{0.0000} &
  \multicolumn{1}{c|}{0.5000} &
  \multicolumn{1}{c|}{0.0000} &
  \multicolumn{1}{c|}{1.0000} &
  \multicolumn{1}{c|}{0.0000} &
  1.0000 \\ \hline
\multicolumn{1}{|c|}{\multirow{4}{*}{\textbf{SIMON32/64}}} &
  \multicolumn{1}{c|}{ResNet} &
  \multicolumn{1}{c|}{0.5002} &
  \multicolumn{1}{c|}{0.5002} &
  \multicolumn{1}{c|}{0.4947} &
  \multicolumn{1}{c|}{0.4974} &
  \multicolumn{1}{c|}{0.5003} &
  \multicolumn{1}{c|}{0.4947} &
  \multicolumn{1}{c|}{0.5057} &
  \multicolumn{1}{c|}{0.4943} &
  0.5053 \\ \cline{2-11} 
\multicolumn{1}{|c|}{} &
  \multicolumn{1}{c|}{CNN} &
  \multicolumn{1}{c|}{0.4993} &
  \multicolumn{1}{c|}{0.4985} &
  \multicolumn{1}{c|}{0.2235} &
  \multicolumn{1}{c|}{0.3086} &
  \multicolumn{1}{c|}{0.4992} &
  \multicolumn{1}{c|}{0.2235} &
  \multicolumn{1}{c|}{0.7750} &
  \multicolumn{1}{c|}{0.3086} &
  0.4992 \\ \cline{2-11} 
\multicolumn{1}{|c|}{} &
  \multicolumn{1}{c|}{LSTM} &
  \multicolumn{1}{c|}{0.5000} &
  \multicolumn{1}{c|}{0.5053} &
  \multicolumn{1}{c|}{0.0009} &
  \multicolumn{1}{c|}{0.0017} &
  \multicolumn{1}{c|}{0.4996} &
  \multicolumn{1}{c|}{0.0008} &
  \multicolumn{1}{c|}{0.9991} &
  \multicolumn{1}{c|}{0.0017} &
  0.4996 \\ \cline{2-11} 
\multicolumn{1}{|c|}{} &
  \multicolumn{1}{c|}{BiLSTM} &
  \multicolumn{1}{c|}{0.5000} &
  \multicolumn{1}{c|}{0.0000} &
  \multicolumn{1}{c|}{0.0000} &
  \multicolumn{1}{c|}{0.0000} &
  \multicolumn{1}{c|}{0.4991} &
  \multicolumn{1}{c|}{0.0000} &
  \multicolumn{1}{c|}{1.0000} &
  \multicolumn{1}{c|}{0.0000} &
  0.4991 \\ \hline
\multicolumn{11}{|c|}{\textbf{Standard Configuration}} \\ \hline
\multicolumn{1}{|c|}{\multirow{4}{*}{\textbf{SPECK32/64}}} &
  \multicolumn{1}{c|}{ResNet} &
  \multicolumn{1}{c|}{0.5000} &
  \multicolumn{1}{c|}{0.5000} &
  \multicolumn{1}{c|}{1.0000} &
  \multicolumn{1}{c|}{0.6667} &
  \multicolumn{1}{c|}{0.5001} &
  \multicolumn{1}{c|}{1.0000} &
  \multicolumn{1}{c|}{0.0000} &
  \multicolumn{1}{c|}{1.0000} &
  0.0000 \\ \cline{2-11} 
\multicolumn{1}{|c|}{} &
  \multicolumn{1}{c|}{CNN} &
  \multicolumn{1}{c|}{0.4997} &
  \multicolumn{1}{c|}{0.4999} &
  \multicolumn{1}{c|}{0.9489} &
  \multicolumn{1}{c|}{0.6548} &
  \multicolumn{1}{c|}{0.4996} &
  \multicolumn{1}{c|}{0.9489} &
  \multicolumn{1}{c|}{0.0505} &
  \multicolumn{1}{c|}{0.9494} &
  0.0511 \\ \cline{2-11} 
\multicolumn{1}{|c|}{} &
  \multicolumn{1}{c|}{LSTM} &
  \multicolumn{1}{c|}{0.5000} &
  \multicolumn{1}{c|}{0.0000} &
  \multicolumn{1}{c|}{0.0000} &
  \multicolumn{1}{c|}{0.0000} &
  \multicolumn{1}{c|}{0.5001} &
  \multicolumn{1}{c|}{0.0000} &
  \multicolumn{1}{c|}{1.0000} &
  \multicolumn{1}{c|}{0.0000} &
  1.0000 \\ \cline{2-11} 
\multicolumn{1}{|c|}{} &
  \multicolumn{1}{c|}{BiLSTM} &
  \multicolumn{1}{c|}{0.4999} &
  \multicolumn{1}{c|}{0.0000} &
  \multicolumn{1}{c|}{0.0000} &
  \multicolumn{1}{c|}{0.0000} &
  \multicolumn{1}{c|}{0.5003} &
  \multicolumn{1}{c|}{0.0000} &
  \multicolumn{1}{c|}{0.9999} &
  \multicolumn{1}{c|}{0.0000} &
  1.0000 \\ \hline
\multicolumn{1}{|c|}{\multirow{4}{*}{\textbf{SIMON32/64}}} &
  \multicolumn{1}{c|}{ResNet} &
  \multicolumn{1}{c|}{0.5000} &
  \multicolumn{1}{c|}{0.5000} &
  \multicolumn{1}{c|}{1.0000} &
  \multicolumn{1}{c|}{0.6667} &
  \multicolumn{1}{c|}{0.5000} &
  \multicolumn{1}{c|}{1.0000} &
  \multicolumn{1}{c|}{0.0000} &
  \multicolumn{1}{c|}{1.0000} &
  0.0000 \\ \cline{2-11} 
\multicolumn{1}{|c|}{} &
  \multicolumn{1}{c|}{CNN} &
  \multicolumn{1}{c|}{0.4999} &
  \multicolumn{1}{c|}{0.4998} &
  \multicolumn{1}{c|}{0.0721} &
  \multicolumn{1}{c|}{0.1260} &
  \multicolumn{1}{c|}{0.5000} &
  \multicolumn{1}{c|}{0.0720} &
  \multicolumn{1}{c|}{0.9278} &
  \multicolumn{1}{c|}{0.1260} &
  0.5000 \\ \cline{2-11} 
\multicolumn{1}{|c|}{} &
  \multicolumn{1}{c|}{LSTM} &
  \multicolumn{1}{c|}{0.5000} &
  \multicolumn{1}{c|}{0.6667} &
  \multicolumn{1}{c|}{0.0000} &
  \multicolumn{1}{c|}{0.0000} &
  \multicolumn{1}{c|}{0.5004} &
  \multicolumn{1}{c|}{0.4999} &
  \multicolumn{1}{c|}{0.5000} &
  \multicolumn{1}{c|}{0.0000} &
  1.0000 \\ \cline{2-11} 
\multicolumn{1}{|c|}{} &
  \multicolumn{1}{c|}{BiLSTM} &
  \multicolumn{1}{c|}{0.5000} &
  \multicolumn{1}{c|}{0.5295} &
  \multicolumn{1}{c|}{0.0005} &
  \multicolumn{1}{c|}{0.0010} &
  \multicolumn{1}{c|}{0.5003} &
  \multicolumn{1}{c|}{0.9996} &
  \multicolumn{1}{c|}{0.0004} &
  \multicolumn{1}{c|}{0.0010} &
  0.5003 \\ \hline
\end{tabular}%
\vspace{-2mm}
\end{table*}

\section{Results}
In our experiment, we evaluated the cryptographic indistinguishability of two lightweight block ciphers, SPECK32/64 and SIMON32/64 using four DL architecture: ResNet, CNN, LSTM, and BiLSTM. We conducted assessments under both round-reduced and standard (full-round) configurations. The classification performance for each configuration is summarized in Table~\ref{tab:dl-results}.

In the round-reduced configuration, both ciphers demonstrate strong indistinguishability. For SPECK32/64, models such as ResNet, LSTM, and BiLSTM achieved near-random performance (accuracy $\approx$50\%, ROC-AUC $\approx$50\%), with zero precision and recall, indicating an inability to differentiate ciphertexts of $\mathcal{P}_{1}$ or $\mathcal{P}_{2}$. The CNN model showed marginal improvement (recall = 0.0356) but remained ineffective, as evidenced by its low F1-score (0.0665). Similarly, for SIMON32/64, ResNet and CNN models exhibited balanced but random-like accuracy ($\approx$50\%), with CNN marginally better at detecting ciphertexts of $\mathcal{P}_{2}$ (recall = 0.2235) but compromised by high false positives (FPR = 0.3086). LSTM and BiLSTM models entirely failed, reinforcing security of the lightweight block ciphers.

In the standard configuration, the results highlight systemic biases rather than meaningful discrimination. For SPECK32/64, ResNet achieved perfect recall (1.0) but trivial precision (0.5), reflecting prediction of all samples as ciphertext belonging to $\xi_{1}$ (or $\mathcal{P}_{1}$), which renders it uninformative. The CNN model exhibited high recall (0.9489) but suffered from severe false positives (FPR = 0.9494), undermining its reliability. For SIMON32/64, ResNet mirrored this behavior, while BiLSTM showed extreme bias (TPR = 0.9996, TNR = 0.0004). Across both ciphers, models like LSTM and BiLSTM consistently failed to generalize, with near-zero recall and precision.

The overarching pattern across configurations is the inability of DL models to surpass random guessing (accuracy and ROC-AUC $\approx$50\%) underscores the cryptographic strength of SPECK32/64 and SIMON32/64 against the considered DL models. While certain models (e.g., CNN for SPECK32/64 in standard configuration) showed skewed metrics,  these reflect algorithmic biases rather than true discriminative capability. The findings confirm that these ciphers maintain strong security evaluated settings.

\section{Discussions}
\label{sec:discussion}

Our experimental results consistently show that ML models fail to surpass random guessing when distinguishing ciphertexts produced by lightweight block ciphers. To better understand these results, we conducted detailed analysis exploring whether models genuinely learns cryptographic patterns or merely memorize overlapping ciphertext samples.

\noindent
\textbf{Analysis of Memorization vs. Generalization: Why ML models Fail to Identify Patterns.} Lightweight block ciphers such as SPECK32/64 and SIMON32/64 generate 32-bit ciphertexts, producing approximately $2^{32}$ (over 4 billion) possible ciphertext outputs for a given plaintext message under full entropy conditions. Exhaustively analyzing such an enormous dataset to detect cryptographic leakage or statistical patterns in computationally prohibitive and practically infeasible due to extensive resources required. Therefore, to conduct computationally manageable evaluation, we intentionally restricted randomness of the initialization vectors (IVs) to 16 bits. Since ciphertext variability directly depends on IV randomness, this restriction reduced the ciphertext space to approximately $2^{16}$ (65,536) unique ciphertexts, creating a controlled yet meaningful experimental scenario to test if ML models genuinely learn or merely memorize ciphertext patterns.

In our primary experiments with SPECK32/64, we observed that when ML models were trained on datasets containing extensive oversampling -- intentional duplication of  ciphertext samples to explicitly test memorization capabilities of ML models -- the models achieved nearly 99\% accuracy. This accuracy reflects memorization of duplicate ciphertext entries rather than genuine generalization. Conversely, when models were trained with only a limited number of unique ciphertext pairs without extensive duplication, accuracy dropped sharply to approximately random guessing ($\approx$50\%) when evaluated on unseen ciphertext pairs. However, these models could still correctly classify ciphertext pairs exactly matching those in the training set, further underscoring the effect of memorization.

To systematically investigation memorization versus genuine generalization in DL models, we performed a detailed analysis on datasets generated with 16-bit IV randomness. Our training dataset comprises of 800,000 ciphertext samples, with an equal split (400,000 each) between ciphertexts of $\mathcal{P}_{1}$ and $\mathcal{P}_{2}$. Within these samples, $\mathcal{P}_{1}$ has 65,395 unique ciphertexts, while $\mathcal{P}_{2}$ has 65,375 unique ciphertexts. The testing dataset contains a total of 100,000 ciphertexts samples, equally distributed between $\mathcal{P}_{1}$ and $\mathcal{P}_{2}$. Specifically, ciphertexts corresponding to $\mathcal{P}_{1}$ include 34,974 unique samples, and those corresponding to $\mathcal{P}_{2}$ include 35,049 unique samples, resulting in combined total of 70,023 unique ciphertexts in the test set.

In our controlled experiment with reduced entropy (16-bits instead of 32-bits), we selected subsets containing 5,000 ciphertext samples per class (10,000 samples in total) from the training dataset. Within this subset, $\mathcal{P}_{1}$ had 4,819 unique ciphertexts, and $\mathcal{P}_{2}$ had 4,815 unique ciphertexts, with 366 redundant samples. Upon examining overlaps between training subset and the complete testing dataset, we identified 5,307 overlapping ciphertext samples. Specifically, 2,659 samples of $\mathcal{P}_{1}$ and 2,684 samples of $\mathcal{P}_{2}$ appeared in both training and testing datasets, constituting approximately 5\% overlap. Such overlaps are crucial, as they directly enable memorization effects by allowing the model to recognize previously encountered samples.

Evaluating the DL model trained on these subsets, we obtained an overall accuracy of about 53.72\%. The cross-validation accuracy was around 52.6\%, slightly above random guessing (50\%), indicating a minimal memorization effect. To further clarify whether the model's performance resulted from genuine generalization or memorization, we conducted detailed sample-by-sample analysis. Among the 70,023 unique ciphertexts samples in the testing dataset, the model correctly classified 53.58\% of them. However, when isolating samples unique only to the testing dataset (thus excluding overlapping training samples), the accuracy sharply dropped to 49.90\%, equivalent to random guessing.

This analysis conclusively demonstrates that ML models fail to identify meaningful cryptographic patterns or statistically exploitable leakage under artificially simplified cryptographic conditions. The observed marginal improvements in accuracy above random chance are entirely due to memorization of overlapping ciphertext samples, rather than genuine generalization by the ML algorithm.

Overall, the inability of state-of-the-art ML models to surpass random guessing underscores not a deficiency of ML techniques, but rather highlights the inherent robustness and strength of cryptographic indistinguishability within lightweight block cipher designs.
\section{Related Work}
\label{lab:related-work} 

\noindent
\textbf{Linear \& Differential Cryptanalysis.} Albrecht et al. \cite{AMGL-12-lr-1} introduced a unified framework that synergistically incorporates various differential cryptanalysis techniques, including standard, truncated, and impossible differentials. These methods are particularly effective in extending the capabilities of known attacks against lightweight block ciphers such as KATAN-32. Following a similar thematic exploration, Dinur et al. \cite{ID-14-imporved-DC-lr-4} and Blondeau et al. \cite{BCBG-11-lr5} refined differential cryptanalysis techniques specifically for a round-reduced version of SPECK, highlighting potential weaknesses of these ciphers under constrained operational conditions. In parallel, Ashur et al. \cite{TADB-16-lr-2} examined the SPECK cipher using linear cryptanalysis, revealing vulnerabilities across various block sizes and demonstrating that linear approximations could be exploited to undermine the cipher's integrity. Complementing these analyses, Biryukov et al. \cite{BAVVYLC-16-lr-3} developed a branch-and-bound method that identifies linear and differential trails in ARX-based ciphers. They specifically applied this approach to enhance cryptanalytic attacks against SPECK. Further studies on the operational constraints of these ciphers also support these findings \cite{FAELSLJW-14-lr6, MRAGL-12-lr7}.

\textbf{ML for Cryptanalysis.} Classical cryptanalysis methods, deeply rooted in the mathematical underpinnings of cryptographic algorithms and ciphertexts, Sabaawi et al. \cite{AS21-cryptanalysis-classical-ciphers} extended these traditional techniques by surveying cryptanalysis implementation on ciphers like Caesar, transposition, and Hill. Simultaneously, Khoirom et al. \cite{Khoirom18} proposed an image encryption scheme based on elliptic curve cryptography and chaotic maps. Their work identified vulnerabilities in the original scheme, leading to an improved version resilient to chosen-plaintext attacks, differential attacks, and statistical attacks, thereby enhancing security and performance in image encryption. This comprehensive exploration spans classical and contemporary approaches, highlighting the evolving landscape of cryptographic techniques for heightened security across diverse applications.

Sikdar et al. \cite{SSKM-22} conducted a survey on recent cryptanalysis techniques, including brute-force attacks, exploring the growing influence of machine learning in cryptographic methods and suggesting future research directions. Verma et al. \cite{VR22} delved into the historical significance of brute-force attacks in cybersecurity, emphasizing their enduring relevance for unauthorized data access. Additionally, Mok et al. \cite{MCJ-19-bruteforce} proposed an intelligent brute-force attack targeting the RSA cryptosystem, simulating and evaluating the effectiveness of their approach in terms of time required for RSA key recovery. Collectively, these works contribute to the understanding and evolution of brute-force cryptanalysis, addressing its challenges and exploring avenues for improved security measures.

While considering side-channel cryptanalysis methods, which focus on the physical characteristics and behaviors of cryptographic devices or implementations, Zhou et al. \cite{ZY-05-sca} provided a comprehensive survey covering methods, techniques, and countermeasures in side-channel attacks, evaluating their feasibility and applicability. In a complementary study, Randolph et al. \cite{RM20-sca} present an in-depth tutorial on power side-channel analysis, spanning the past two decades. The study elucidates fundamental concepts and practical applications of various attacks, such as Simple Power Analysis (SPA), Differential Power Analysis (DPA), Template Attacks (TA), Correlation Power Analysis (CPA), Mutual Information Analysis (MIA), and Test Vector Leakage Assessment (TVLA), along with the underlying theories. Additionally, the introduction of test statistics as a measure of confidence in detecting side-channel leakage adds depth to these analyses.

Mehmood et al. \cite{MZ23} conducted a comprehensive evaluation of distinguishability on the ciphertexts of AES-128 cipher in CBC and ECB modes. Their methodology involved employing Support Vector Machine, k-Nearest Neighbours, and Random Forest Classifiers trained on the frequency distribution of characters in the ciphertexts. The results underscored the susceptibility of the ECB mode, thereby emphasizing the need for robust encryption techniques. Building upon this foundation, Hu et al. \cite{Hu19} explored by applying Random Forest classifiers to diverse block ciphers, reinforcing the vulnerability of the ECB mode. These studies not only showcase the evolving landscape of machine learning-based cryptanalysis but also highlight its role in ensuring the resilience of cryptographic algorithms.

Xiao et al. \cite{XY-19-cryptanalysis-neural} significantly contributed to the field of neural network (NN) based cryptanalysis by introducing a novel approach that not only focuses on the development of neural distinguishers but also emphasizes metrics for efficacy assessment. Their framework, applied to Cyber-Physical Systems (CPS) ciphers, adds depth to the understanding of NN-based cryptanalysis.

In summary, while the reviewed literature presents a comprehensive understanding of various cryptanalysis methods, it is noteworthy that the majority of the approaches explores differential attacks, statistical attacks, chosen-plaintext attacks, etc. In contrast to prior research, our work addresses a critical gap in the literature and providing a more comprehensive evaluation of cryptographic indistinguishability of lightweight block ciphers.
\section{Conclusion}
\label{sec:conclusion}

In this research, we introduced a machine learning-based framework, \attackname, designed specifically to assess the cryptographic indistinguishability of SPECK32/64 and SIMON32/64 lightweight block ciphers. Our investigation utilized various state-of-the-art deep learning architectures to assess these ciphers using machine learning. 

Our results show that deep learning models fail to surpass random guessing accuracy ($\approx$50\%) in distinguishing ciphertexts of two plaintext messages $\mathcal{P}_{1}$, and $\mathcal{P}_{2}$ encrypted using same key. Our analysis for memorization versus generalization evaluations, further revealed that ML models were memorizing ciphertext samples rather than genuinely learning cryptographic patterns. Even in artificially simplified cryptographic environments with deliberately reduced entropy, ML algorithms exhibited no ability to generalize beyond memorized ciphertexts.

These results provide strong empirical evidence that current ML algorithms, despite their advanced pattern-recognition capabilities, remain ineffective in compromising the indistinguishability property of even lightweight cryptographic algorithms. Future research directions could focus on exploring emerging cryptographic algorithms, advanced ML architectures, or quantum-inspired ML methods, to monitor and validate cryptographic resilience.

\bibliographystyle{ieeetr}
\bibliography{references}

\end{document}